\documentclass[]{pasj02} 
\usepackage[switch,mathlines]{lineno} 
\usepackage{natbib} 
\usepackage{lscape}
\usepackage{url}

\jyear{2024}
\Received{2024/08/20}
\Accepted{2024/11/16}

\newcommand{\oi}{[O {\sc i}]}
\newcommand{\oiii}{[O {\sc iii}]}
\newcommand{\ci}{[C {\sc i}]}
\newcommand{\cii}{[C {\sc ii}]}
\newcommand{\nii}{[N {\sc ii}]}
\newcommand{\siiii}{[Si {\sc iii}]}

\begin{document} 

\title{Detection of the \oi\ $63$ \micron\ emission line from the $z = 6.04$ quasar J2054-0005
}

\author{
 Nozomi \textsc{Ishii},\altaffilmark{1,2}
 Takuya \textsc{Hashimoto},\altaffilmark{1,3}\orcid{0000-0002-0898-4038}\altemailmark \email{hashimoto.takuya.ga@u.tsukuba.ac.jp}
 Carl \textsc{Ferkinhoff},\altaffilmark{4}\orcid{0000-0001-6266-0213}
 Matus \textsc{Rybak},\altaffilmark{5,6,7}\orcid{0000-0002-1383-0746}
 Akio K. \textsc{Inoue},\altaffilmark{8,9}\orcid{0000-0002-7779-8677}
 Tomonari \textsc{Michiyama},\altaffilmark{10}\orcid{0000-0003-2475-7983}
 Darko \textsc{Donevski},\altaffilmark{11,12}\orcid{0000-0001-5341-2162}
 Seiji \textsc{Fujimoto},\altaffilmark{13}\orcid{0000-0001-7201-5066}
 Dragan \textsc{Salak},\altaffilmark{14,15}\orcid{0000-0002-3848-1757}
 Nario \textsc{Kuno},\altaffilmark{1,3}\orcid{0000-0002-1234-8229}
 Hiroshi \textsc{Matsuo},\altaffilmark{16,17}\orcid{0000-0003-3278-2484}
 Ken \textsc{Mawatari},\altaffilmark{1,3,9}\orcid{0000-0003-4985-0201}
 Yoichi \textsc{Tamura},\altaffilmark{18}\orcid{0000-0003-4807-8117}
 Takuma \textsc{Izumi},\altaffilmark{16,17}\orcid{0000-0001-9452-0813}
 Tohru \textsc{Nagao},\altaffilmark{19}\orcid{0000-0002-7402-5441}
 Yurina \textsc{Nakazato},\altaffilmark{20}\orcid{0000-0002-0984-7713}
 Wataru \textsc{Osone},\altaffilmark{1}
 Yuma \textsc{Sugahara},\altaffilmark{9,15}\orcid{0000-0001-6958-7856}
 Mitsutaka \textsc{Usui},\altaffilmark{1}
 Koki \textsc{Wakasugi},\altaffilmark{1}
 Hidenobu \textsc{Yajima},\altaffilmark{1,21}\orcid{0000-0002-1319-3433}
 Tom J. L. C. \textsc{Bakx},\altaffilmark{22}\orcid{0000-0002-5268-2221}
 Yoshinobu \textsc{Fudamoto},\altaffilmark{23}\orcid{0000-0001-7440-8832}
 Romain A. \textsc{Meyer},\altaffilmark{24}\orcid{0000-0001-5492-4522}
 Fabian \textsc{Walter},\altaffilmark{25}\orcid{0000-0003-4793-7880}
 and 
 Naoki \textsc{Yoshida}\altaffilmark{20,26,27}\orcid{0000-0001-7925-238X}
 }

\altaffiltext{1}{
Graduate School of Pure and Applied Sciences, University of Tsukuba, 1-1-1 Tennodai,
Tsukuba, Ibaraki 305-8571, Japan
}
\altaffiltext{2}{
Systems Engineering Consultants Co.,LTD. (SEC)
Setagaya Business Square
4-10-1 Yoga, Setagaya-ku, Tokyo 158-0097, Japan
}
\altaffiltext{3}{
Tomonaga Center for the History of the Universe, University of Tsukuba, 1-1-1 Tennodai,
Tsukuba, Ibaraki 305-8571, Japan
}
\altaffiltext{4}{
Department of Physics, Winona State University, Winona, MN 55987, USA
}
\altaffiltext{5}{
Faculty of Electrical Engineering, Mathematics and Computer Science, Delft University of Technology, Mekelweg 4, 2628 CD Delft, The Netherlands
}
\altaffiltext{6}{
Leiden Observatory, Leiden University, P.O. Box 9513, 2300 RA Leiden, The Netherlands
}
\altaffiltext{7}{
SRON - Netherlands Institute for Space Research, Niels Bohrweg 4, 2333 CA Leiden, The Netherlands
}
\altaffiltext{8}{
Department of Physics, School of Advanced Science and Engineering, Faculty of Science and Engineering, Waseda University, 3-4-1 Okubo, Shinjuku, Tokyo 169-8555, Japan 
}
\altaffiltext{9}{
Waseda Research Institute for Science and Engineering, Faculty of Science and Engineering, Waseda University, 3-4-1 Okubo, Shinjuku, Tokyo 169-8555, Japan
}
\altaffiltext{10}{
Faculty of Information Science, Shunan University, 843-4-2 Gakuendai, Shunanshi, Yamaguchi 745-8566, Japan
}
\altaffiltext{11}{
National Center for Nuclear Research, Pasteura 7, 02-093 Warsaw, Poland
}
\altaffiltext{12}{
SISSA, Via Bonomea 265, 34136 Trieste, Italy
}
\altaffiltext{13}{
Department of Astronomy, The University of Texas at Austin, Austin, TX 78712, USA
}
\altaffiltext{14}{
Institute for the Advancement of Higher Education, Hokkaido University, Kita 17 Nishi 8, Kita-ku, Sapporo, Hokkaido 060-0817, Japan
}
\altaffiltext{15}{
Department of Cosmosciences, Graduate School of Science, Hokkaido University, Kita 10 Nishi 8, Kita-ku, Sapporo, Hokkaido 060-0810, Japan
}
\altaffiltext{16}{
National Astronomical Observatory of Japan, 2-21-1 Osawa, Mitaka, Tokyo 181-8588, Japan
}
\altaffiltext{17}{
The Graduate University for Advanced Studies (SOKENDAI), 2-21-1 Osawa, Mitaka, Tokyo 181-8588, Japan
}
\altaffiltext{18}{
Department of Physics, Graduate School of Science, Nagoya University Furo, Chikusa, Nagoya, Aichi 464-8602, Japan
}
\altaffiltext{19}{
Research Center for Space and Cosmic Evolution, Ehime University, 2-5 Bunkyo-cho, Matsuyama, Ehime 790-8577, Japan
}
\altaffiltext{20}{
Department of Physics, The University of Tokyo, 7-3-1 Hongo, Bunkyo, Tokyo 113-0033, Japan
}
\altaffiltext{21}{
Center for Computational Sciences, University of Tsukuba, 1-1-1 Tennodai, Tsukuba, Ibaraki 305-8577, Japan
}
\altaffiltext{22}{
Department of Earth and Space Sciences, Chalmers University of Technology, Onsala Observatory, 439 94 Onsala, Sweden
}
\altaffiltext{23}{
Center for Frontier Science, Chiba University, 1-33 Yayoi-cho, Inage-ku, Chiba 263-8522, Japan
}
\altaffiltext{24}{
Department of Astronomy, University of Geneva, Chemin Pegasi 51, 1290 Versoix, Switzerland
}
\altaffiltext{25}{
Max-Planck-Institut für Astronomie, Königstuhl 17, D-69117 Heidelberg, Germany
}
\altaffiltext{26}{
Kavli Institute for the Physics and Mathematics of the Universe (WPI), UT Institute for Advanced Study, The University of Tokyo, Kashiwa, Chiba 277-8583, Japan
}
\altaffiltext{27}{
Research Center for the Early Universe, School of Science, The University of Tokyo, 7-3-1 Hongo, Bunkyo, Tokyo 113-0033, Japan
}


\maketitle

\begin{abstract}
We report the highest-redshift detection of \oi\ 63 \micron\ from a luminous quasar, J2054-0005, at $z=6.04$ based on the Atacama Large Millimeter/sub-millimeter Array Band 9 observations. The \oi\ 63 \micron\ line luminosity is $(4.5\pm1.5) \times 10^{9}~L_{\rm \odot}$, corresponding to the \oi\ 63 \micron-to-far-infrared luminosity ratio of $\approx 6.7\times10^{-4}$, which is consistent with the value obtained in the local universe. Remarkably, \oi\ 63 \micron\ is as bright as \cii\ 158 \micron, resulting in the \oi-to-\cii\ line luminosity ratio of $1.3\pm0.5$. Based on a careful comparison of the luminosity ratios of \oi\ 63 \micron, \cii\ 158 \micron, and dust continuum emission to models of photo-dissociation regions, we find that J2054-0005 has a gas density log($n_{\rm H}$/cm$^{-3}$)$=3.7\pm0.3$ and an incident far-ultraviolet radiation field of log($G/G_{\rm 0}$)~$= 3.0\pm0.1$, showing that \oi\ 63 \micron\ serves as an important coolant of the dense and warm gas in J2054-0005. A close examination of the \oi\ and \cii\ line profiles suggests that the \oi\ line may be partially self-absorbed, however deeper observations are needed to verify this conclusion. Regardless, the gas density and incident radiation field are in a broad agreement with the values obtained in nearby star-forming galaxies and  objects with \oi\ 63 \micron\ observations at $z=1-3$ with the Herschel Space Observatory. These results demonstrate the power of ALMA high-frequency observations targeting \oi\ 63 \micron\ to examine the properties of photo-dissociation regions in high-redshift galaxies. \end{abstract}


\section{Introduction}
\label{sec:intro}

Quasars are powered by supermassive black holes (SMBHs) with masses of $\sim 10^{8-10} M_{\rm \odot}$ (e.g., \citealt{inayoshi2020} and references therein). In the local Universe, there is a well-known correlation between the central black hole mass and the bulge mass (e.g., \citealt{kormendy2013}). Given the coevolution of SMBHs and their host galaxies, understanding  detailed properties of the host galaxies in the early universe is a key to understand the formation and evolution of SMBHs (e.g., \citealt{valiante2017}). 

The interstellar medium (ISM) of high-redshift ($z$) quasar host galaxies is often studied with the brightest far-infrared (FIR) fine structure line, \cii\ 158 \micron, as well as with carbon monoxide (CO) emission lines, and dust continuum emission (e.g., \citealt{venemans2018, decarli2018,izumi2019,decarli2020, pensabene2024, tripodi2024}). These studies have shown that the quasar host galaxies can be characterized by high star-formation rates (SFR~$\sim50-3000$~$M_{\rm \odot}$~yr$^{-1}$), molecular gas masses ($\sim 10^{10}$~$M_{\rm \odot}$), and dust masses ($\sim 10^{7-9}$~$M_{\rm \odot}$). 

The \oi\ $^{3}P_{2} \rightarrow ^{3}P_{1}$ line at the rest-frame wavelength of $63$ \micron\ (rest-frame frequency of $4744.77749$ GHz), hereafter \oi\ 63 \micron, has a critical density of $n_{\rm H, crit.} \sim 2.5 \times 10^{5}$~cm$^{-3}$ at the temperature of 100 K and emitting level with an equivalent temperature of $T_{\rm *} = \Delta E/k = 228$ K above the ground (\citealt{draine2011}). Compared to the critical density of \cii\ 158 \micron, $n_{\rm H, crit.} \sim 3 \times 10^{3}$~cm$^{-3}$ at the temperature of 100 K, and the equivalent temperature $T_{\rm *} = \Delta E/k = 91.25$ K, the \oi\ 63 \micron\ line traces high-density, high-temperature neutral gas  (\citealt{kaufman1999, kaufman2006, narayanan2017}). Thus, combinations of \oi\ 63 \micron, \cii\ 158 \micron, and dust continuum data allow us to estimate the gas density and temperature in photo-dissociated regions (PDRs). Due to the high equivalent temperature, \oi\ 63 \micron\ also has an advantage that it is less affected by the cosmic microwave background radiation, which makes it a reliable tracer of neutral ISM at high redshift.

Initial campaigns for the \oi\ 63 \micron\ line observations include the NASA Lear jet telescope system targeting the Orion nebula (M42) and Omega nebular (M17) (\citealt{melnick1979}) and a balloon-borne telescope targeting the Orion nebula (\citealt{furniss1983}). 
The \oi\ 63 \micron\ line was observed in external galaxies with the Infarad Space Observatory (ISO) (e.g., \citealt{malhotra2001, brauher2008}) and the Herschel Space Observatory (hereafter Herschel) (e.g., \citealt{farrah2013, madden2013, cormier2015, diaz-santos2017, herrera-camus2018}) in metal-rich spiral galaxies, ultra-luminous infrared galaxies (ULIRGs), and blue compact dwarf galaxies. 
These observations showed that \oi\ 63 \micron\ is a dominant coolant in dense PDRs, with a typical \oi-to-FIR luminosity ratio $L_{\rm [OI]63\micron}/L_{\rm FIR}$ $\approx 10^{-4} - 10^{-3}$, where $L_{\rm FIR}$ is integrated over $42-122~\micron$.

The \oi\ 63 \micron\ line observations at $z\sim1-3$ were conducted by Herschel, mainly in gravitationally-lensed sub-millimeter galaxies (SMGs) (e.g., \citealt{sturm2010, coppin2012, brisbin2015, wardlow2017, zhang2018, wagg2020}) and in the Cloverleaf quasar at $z=2.46$ (\citealt{uzgil2016}). 
The detections unveiled enhanced luminosity ratios, $L_{\rm [OI]63\micron}/L_{\rm FIR}$ ranging from $\approx 10^{-3}$ to $10^{-2}$, but still with a limited number of detections ($N\sim 15$).
On the other hand, a stacking of the non-detections of individual SMGs in \cite{wardlow2017} resulted in a detection of \oi\ 63 \micron, yielding $L_{\rm [OI]63\micron}/L_{\rm FIR} = (3.6 \pm 1.2) \times 10^{-4}$, showing a diversity in the ISM properties of high-$z$ galaxies. 

At $z \sim 4-7$, the \oi\ 63 \micron\ line can be observed from the ground with e.g., the Atacama Large Millimeter/sub-millimeter Array (ALMA) in Band 9 - 10. 
The \oi\ 63 \micron\ line observation at $z>4$ has been first reported by \cite{rybak2020}. The authors observed a gravitationally-lensed SMG, G09 83808, at $z=6.027$ with the Atacama Pathfinder EXperiment (APEX) 12-m telescope. Although a $5\sigma$ \oi\ 63 \micron\ line was initially reported, subsequent follow-up observations with ALMA Band 9 did not confirm the line (\citealt{rybak2023}). 
\cite{litke2022} targeted \oi\ 63 \micron\ along with other FIR lines in a gravitationally-lensed SMG, SPT 0346-52, at $z=5.656$ with ALMA Band 9, resulting in a non-detection of \oi\ 63 \micron. 
The first detection of \oi\ 63 \micron\ at $z>4$ was recently made with ALMA Band 10 in a non-lensed hyper-luminous ($L_{\rm FIR} \sim 3.5\times10^{13}~L_{\odot}$) active galactic nucleus (AGN), W2246-0526, at $z=4.6$ (\citealt{fernandez_aranda2024}). 
Clearly, the number of \oi\ 63 \micron\ observations at $z>4$ is still scarce. 
A contributing factor to this scarcity may be that \oi\ 63 \micron\ emission is typically optically thick (e.g., \citealt{goldsmith2021}), unlike most of far-IR fine-structure lines like \cii\ 158 \micron. As such, \oi\ 63 \micron\ emission is detected only from front faces of emitting clouds. Additionally, if sufficient neutral oxygen is along the line-of-sight in a galaxy, the \oi\ 63 \micron\ line can be readily self-absorbed. For example, self-absorption is seen in a nearby large molecular cloud DR 21 (\citealt{poglitsch1996}), a diffused nebula NGC 6334 (\citealt{kraemer1998}), and a highly obscured {\sc H ii} region Sgr B2 (\citealt{baluteau1997}) with massive star formation activity. In extra-galactic sources, nearly half of the local ULIRGS in \cite{rosenberg2015} show significant missing flux with several sources show the 63 \micron\  line nearly completely absorbed out (see also \citealt{gonzalez-alfonso2012}). Determining if \oi\ 63 \micron\  emission is optically thick requires comparing the line to the optically thin \oi\ 145 \micron\ line, while assessing self-absorption requires a significant signal-to-noise ratio in the spectrum to identify absorption features. \cite{fernandez_aranda2024} detect both the \oi\ 63 and 145 \micron\ lines, and show their line ratios suggest the \oi\ 63 \micron\ line emission is indeed optically thick. However, the data were not sensitive enough to identify the narrow absorption feature in W2246-0526. 

In this paper, we report a new detection of \oi\ $63$ \micron\ in a distant quasar, SDSS J2054-0005, at $z = 6.0$ based on ALMA Band 9 archival data (\S \ref{sec:sample_data}). To date, this is the most distant detection of \oi\ 63 \micron. We present our calculations on the \oi\ 63 \micron\ luminosity and its luminosity ratios against $L_{\rm FIR}$ and the \cii\ 158 \micron\ luminosity in \S \ref{sec:results}. In \S \ref{sec:discussion}, we perform PDR modeling to constrain the neutral gas properties. We summarize our results in \S \ref{sec:summary}. 

Through out this paper, we assume a $\Lambda$ cold dark matter cosmology with $\Omega_{m} = 0.272$, $\Omega_{\Lambda} = 0.728$, and $H_{0} = 70.4$~km~s$^{-1}$~Mpc$^{-1}$ (\citealt{komatsu2011}). At the
redshift of the source ($z = 6.04$), the age of the universe is 0.946 Gyr and an angular size of $1''$ corresponds to a proper distance of 5.85 kpc. 
The solar luminosity is $L_{\odot} = 3.839 \times 10^{33}$~erg~s$^{-1}$.

\section{Observations and Data}
\label{sec:sample_data}

\begin{table*}[t]
  \caption{ALMA Band 9 Data}
  \centering
  \begin{tabular}{cccc} \hline
    Data & Achieved sensitivity in r.m.s. & Beam size FWHM & Beam PA \\ 
     & (mJy/beam) & $\rm{(arcsec \times arcsec)}$ & (degree) \\ \hline
    Dust continuum & 0.55 & $0.64 \times 0.56$ & 72 \\ 
   {\rm [O$\rm I$] 63 $\mu$m cube} (50 km~s$^{-1}$ bin) & 3.68 & $0.65 \times 0.56$ & 83 \\ \hline
  \end{tabular}
  \label{tab1}
\end{table*}

\subsection{The target: J2054-0005}
\label{subsec:sample}

The quasar J2054-0005 was originally discovered in the Sloan Digital Sky Survey (SDSS) (\citealt{jiang2008}), and later detected in various emission lines of \cii\ $158$ \micron\ (e.g., \citealt{wang2013, venemans2020}), \oiii\ $88$ \micron\ (\citealt{hashimoto2019}),  CO(2-1) and CO(6-5) (\citealt{shao2019}), CO(7-6) (\citealt{decarli2020}), as well as in the OH $119$ \micron\ absorption and emission (\citealt{salak2024}). The redshift determined from these lines is $6.0391 \pm 0.0002$. J2054-0005 was also observed in \ci(2-1), however, it resulted in a non-detection (\citealt{decarli2020}).

J2054-0005 is a non-lensed quasar, and has a bolometric luminosity of $\approx 3 \times 10^{13}~L_{\odot}$ (\citealt{wang2013, farina2022}). The black hole mass is estimated to be $\sim (1.5-10.2) \times 10^{9}~M_{\odot}$ based on the single-epoch technique with an Eddington ratio of $\lambda_{\rm Edd} \sim 0.1-0.7$ (\citealt{farina2022}). Based on a fit to multi-wavelength dust continuum data, J2054-0005 has a total infrared luminosity integrated over $8-1000$ \micron, $L_{\rm TIR}$, $\approx 1 \times 10^{13}~L_{\odot}$ \citep{wang2013, hashimoto2019, tripodi2024}, making it one of the brightest quasars at $z\gtrsim6$. The SFR estimated from $L_{\rm TIR}$ is $\approx 800~M_{\odot}$~yr$^{-1}$ after a correction of the AGN contribution \citep{salak2024}.

\begin{figure*}[hbtp]
\includegraphics[width=17cm]{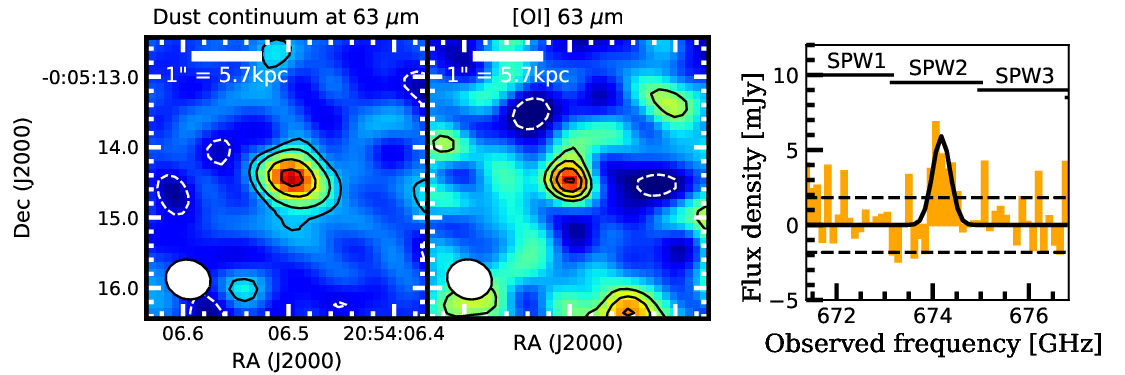}
\caption
{
The dust continuum map at $\approx 63$ \micron\ (left), \oi\ 63 \micron\ integrated intensity map (middle), and continuum-subtracted \oi\ 63 \micron\ spectrum (right). In the left and middle panels, the ellipse at the lower left corner indicates the synthesized beam size of ALMA. The dust continuum contours are drawn at $\pm$(2, 4, 8, 12) $\times \sigma$, where $\sigma = 0.55$~mJy~beam$^{-1}$. The \oi\ 63 \micron\ line contours are drawn at $\pm$(2, 3, 4, 5) $\times \sigma$, where $\sigma = 0.42$~Jy~beam$^{-1}$~km~s$^{-1}$. Negative and positive contours are shown by the white dashed and black solid lines, respectively. In the right panel, the continuum-subtracted \oi\ 63 \micron\ spectrum is extracted from the region with $>2\sigma$ detection in the integrated intensity map. The black solid line is the best-fit Gaussian for the line, whereas the black dashed lines show the typical $\pm1\sigma$ noise level measured from the spectrum. The upper horizontal bars indicate the frequency coverages of SPWs1-3. 
}
\label{fig1}
\end{figure*}

\subsection{ALMA Band 9 Data}

ALMA Band 9 observation was carried out on 2017 March 22 as part of the ALMA Cycle 4 program (Project ID: 2016.1.01063.S, PI: C. Ferkinhoff). The antennas of the 12 m array observed a single field centered at ($\alpha_{\rm ICRS}$, $\delta_{\rm ICRS}$) = ($20^h 54^m 06\fs 490, -00\arcdeg 05' 14\farcs80$). Forty-two antennas were used, with baseline lengths ranging from $15.1 - 278.9$~m. This resulted in a maximum recoverable scale of $4\farcs2$ and a field-of-view of $8\farcs6$. The on-source integration time was $39$ minutes. 

Four spectral windows were set at central frequencies of $672.3$, $674.1$, $676.0$, and $677.8$ GHz, referred to as SPW1, SPW2, SPW3, and SPW4, respectively. SPW2 observed the \oi\ 63 \micron\ line, whereas the other three SPWs observed dust continuum emission. 
The quasar J1924-2914 and J2025-0735 was used for the bandpass and complex gain calibration, respectively. The flux was scaled using Titan, yielding a calibration uncertainty of less than 20\%\ in ALMA Band 9, according to the ALMA Cycle 4 Proposer's Guide\footnote{https://almascience.nao.ac.jp/proposing/documents-and-tools/cycle4/alma-proposers-guide}.

The data were reduced with the Common Astronomy Software Applications (CASA, \citealt{mcmullin2022}) version 4.7.0. The CASA {\tt tclean} task was used for imaging, with a natural weighting to optimize the point-source sensitivity. Table \ref{tab1} summarizes the resulting resolution and sensitivity of the data. 

Continuum maps were created using all channels except for SPW2 that included \oi\ 63 \micron. The synthesized beam has a size of $0\farcs64 \times 0\farcs56$ in the FWHM and a positional angle of $72^{\circ}$. The rms of the map is $0.55~\rm{mJy~beam^{-1}}$.

To create line cubes in SPW2, the CASA task {\tt imcontsub}\footnote{The phase-tracking center of the ALMA Band 9 observation is slightly offset from the position of J2054-0005, ($\alpha_{\rm ICRS}$, $\delta_{\rm ICRS}$) = ($20^h 54^m 06\fs 503, -00\arcdeg 05' 14\farcs43$) in \cite{hashimoto2019}. In this case, {\tt imcontsub} optimally subtracts the continuum compared with the CASA task {\tt uvcontsub} (see e.g., \citealt{kaasinen2023} for the choice of {\tt imcontsub} due to the same issue).} was first applied to subtract the continuum emission. We then created cubes with a velocity width of $50$~km~s$^{-1}$ without {\it uv}-tapering. The cube has a synthesized beam size of $0\farcs65 \times 0\farcs56$ and a positional angle of $83^{\circ}$.  A typical rms sensitivity of the cube is $3.68$ mJy~beam$^{-1}$ per $50$~km~s$^{-1}$ bin. 

\section{Results}
\label{sec:results}

\subsection{$63$-\micron\ continuum emission}
\label{sec:result_dust}

The data probe the dust continuum emission at a rest-frame wavelength of $\lambda_{\rm rest} \approx 63$ \micron. The dust continuum emission is detected at a peak significance of $13\sigma$, as shown in the left panel of figure \ref{fig1}. With the CASA task {\tt imfit}, we obtain the flux density of $S_{\rm \nu, 63 \mu m} = 9.6 \pm 1.0$~mJy (table \ref{tab2}). The peak position is at ($\alpha_{\rm ICRS}$, $\delta_{\rm ICRS}$) = ($20^h 54^m 06\fs 495, -00\arcdeg 05' 14\farcs46$), well consistent with previous dust continuum positions (e.g., \citealt{wang2013, hashimoto2019, salak2024}). The beam-deconvolved size of the continuum-emitting region is $(0\farcs42 \pm 0\farcs13) \times (0\farcs23 \pm 0\farcs09)$, consistent with previous measurements within $2\sigma$ uncertainties: $(0\farcs23 \pm 0\farcs01) \times (0\farcs15 \pm 0\farcs02)$ at $\lambda_{\rm rest} \approx87$ \micron\ \citep{hashimoto2019} and $(0\farcs27 \pm 0\farcs03) \times (0\farcs26 \pm 0\farcs03)$ at $\lambda_{\rm rest} \approx160$ \micron\ \citep{wang2013}. Finally, we note that the results on ALMA Band 9 dust continuum are recently reported in \cite{tripodi2024} based on the same data set. Our results are consistent with their results in the size and flux density measurements. Hereafter, we focus on \oi\ 63 \micron\ in this study. 

\begin{table}[t]
  \caption{Summary of Observational Results}
  \centering
  \begin{tabular}{c c} \hline 
    & \oi\ 63 \micron \\
    \hline 
    $z$ & $6.0385 \pm 0.0005$ \\ 
    FWHM (km~s$^{-1}$) & $192 \pm 49$ \\ 
    $S_{\rm [OI]63}\Delta v$ (Jy~km~s$^{-1}$) & $1.81 \pm 0.61$ \\
    $L_{\rm [OI]63}$ ($L_{\rm \odot}$) & $(4.5\pm1.5)\times10^{9}$ \\ 
    size (beam-convolved) & ($0\farcs58 \pm 0\farcs11$) $\times$ ($0\farcs51 \pm 0\farcs08$) \\
    size (beam-deconvolved) & $< (0\farcs42 \times 0\farcs23)$  \\
    \hline 
    & dust continuum \\
    \hline 
    $S_{\rm \nu, 63 \micron}$ (mJy) & $9.6\pm1.0$ \\
    size (beam-convolved) & ($0\farcs76 \pm 0\farcs02$) $\times$ ($0\farcs61 \pm 0\farcs02$) \\
    size (beam-deconvolved) & ($0\farcs42 \pm 0\farcs13$) $\times$ ($0\farcs23 \pm 0\farcs09$) \\
    \hline
  \end{tabular}
  \label{tab2}
\end{table}

\subsection{\oi\ $63$ \micron\ }
\label{subsec:result_OI}

The middle panel of figure \ref{fig1} shows the \oi\ 63 \micron\ intensity map (i.e., moment 0 map) integrated over the frequency range of $673.96 - 674.44$~GHz, corresponding to a velocity width of $250$~km~s$^{-1}$ ($\approx$ FWHMs of \cii\ 158 \micron\ and \oiii\ 88 \micron) centered at the observed redshift of $6.0391$. The noise level of the map is $\sigma = 0.42$ Jy beam$^{-1}$ km s$^{-1}$. The line is detected at the peak significance level of $5.0\sigma$. The peak position is at ($\alpha_{\rm ICRS}$, $\delta_{\rm ICRS}$) = ($20^h 54^m 06\fs 501, -00\arcdeg 05' 14\farcs42$), well consistent with those for other emission lines (see \citealt{salak2024} for a detailed comparison on the peak positions of emission lines and the optical continuum emission from the quasar). 

The right panel of figure \ref{fig1} shows the continuum-subtracted spectrum extracted from an aperture that encloses the $2\sigma$-region in the integrated intensity map. We apply a Gaussian profile to the spectrum with the {\it scipy }{\tt curve\_fit} function, where we use the SPWs of 1, 2, and 3. To obtain uncertainties on fitting parameters, the noise level is measured from the frequency ranges in [671.3:673.5] GHz and [674.6:676.8] GHz. We obtain the line FWHM of $192\pm49$~km~s$^{-1}$ and $z=6.0385\pm0.0005$. We measure the line flux with the CASA task {\tt imfit} assuming a 2D Gaussian profile for the intensity map integrated over a velocity width of 500 km~s$^{-1}$ ($\approx 2.6\times$FWHM), resulting in $1.81 \pm 0.61$~Jy~km~s$^{-1}$. With Equation (\ref{equ:lumi}) (\citealt{solomon1992}), the luminosity is estimated to be $(4.53 \pm 1.52) \times 10^{9}L_{\odot}$, 
\begin{eqnarray}
  L_{\rm line} = 1.04 \times 10^{-3} \left(\frac{S_{\rm line}\Delta v}{\rm Jy\, km\, s^{-1}}\right) \left(\frac{D_{\rm L}}{\rm Mpc}\right)^{2} \left(\frac{\nu_{\rm obs}}{\rm GHz}\right)L_{\odot},
  \label{equ:lumi}
\end{eqnarray}
where ${S_{\rm line} \Delta v}$, $D_{\rm L}$, and $\nu_{\rm obs}$ are the line flux, luminosity distance, and observed frequency, respectively. The \oi\ 63 \micron-emitting region is not spatially resolved. Under a reasonable assumption that the region has a smaller size than the continuum-emitting region (figure \ref{fig1}), we obtain the upper limit on the size to be $0\farcs42\times0\farcs23$.
Table \ref{tab2} summarizes the observational results.

\subsection{Luminosity Ratios}
\label{subsec:result_luminosityRatio}

\begin{table*}[h]
  \caption{Summary of the luminosities in J2054-0005}
  \centering
{
  \begin{tabular}{cccccc} \hline
  Lines & FWHM & Luminosity & References \\ 
   & (km~s$^{-1}$) & ($L_{\odot}$) & \\ \hline
   \oi\ $63$ \micron\  & $192\pm49$ & $(4.5 \pm 1.5) \times 10^{9}$ & This Study \\
   \cii\ $158$ \micron\  & $243\pm 10$ & $(3.4 \pm 0.5) \times 10^{9}$ & \citet{wang2013} \\ 
   \oiii\ $88$ \micron\  & $282 \pm 17$ & $(6.8 \pm 0.6) \times 10^{9}$ & \citet{hashimoto2019} \\
  \ci(2-1) & - & $< 1.1 \times 10^{8}$ & \citet{decarli2020} \\ 
  \rm{FIR} & - & $(6.8 \pm 0.2) \times 10^{12}$ & \citet{salak2024} \\  \hline
  \end{tabular}
  }
  \label{tab3}
\end{table*}

\begin{figure*}[hbtp]
\includegraphics[width=8cm]{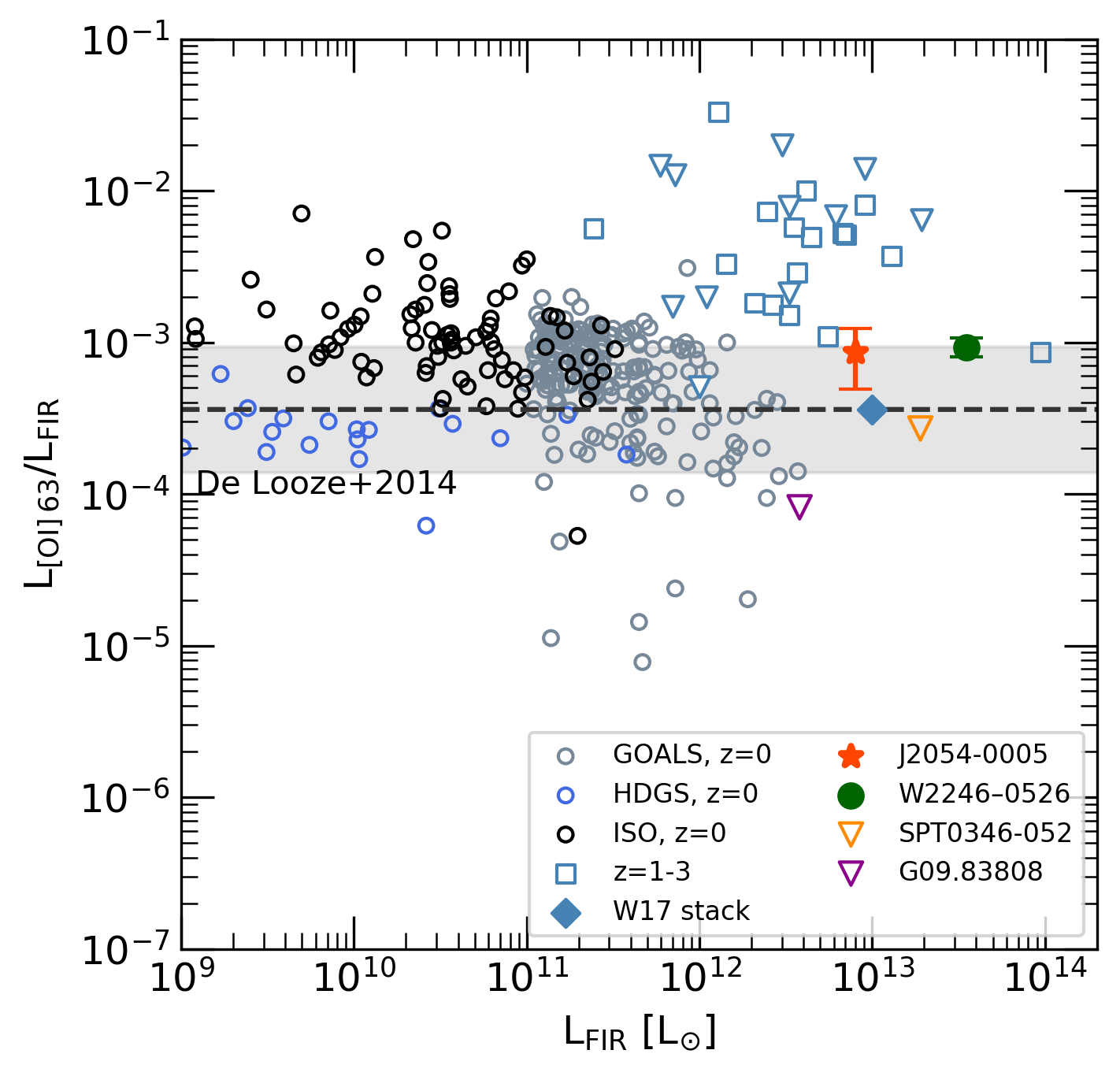}
\hfill
\includegraphics[width=8cm]{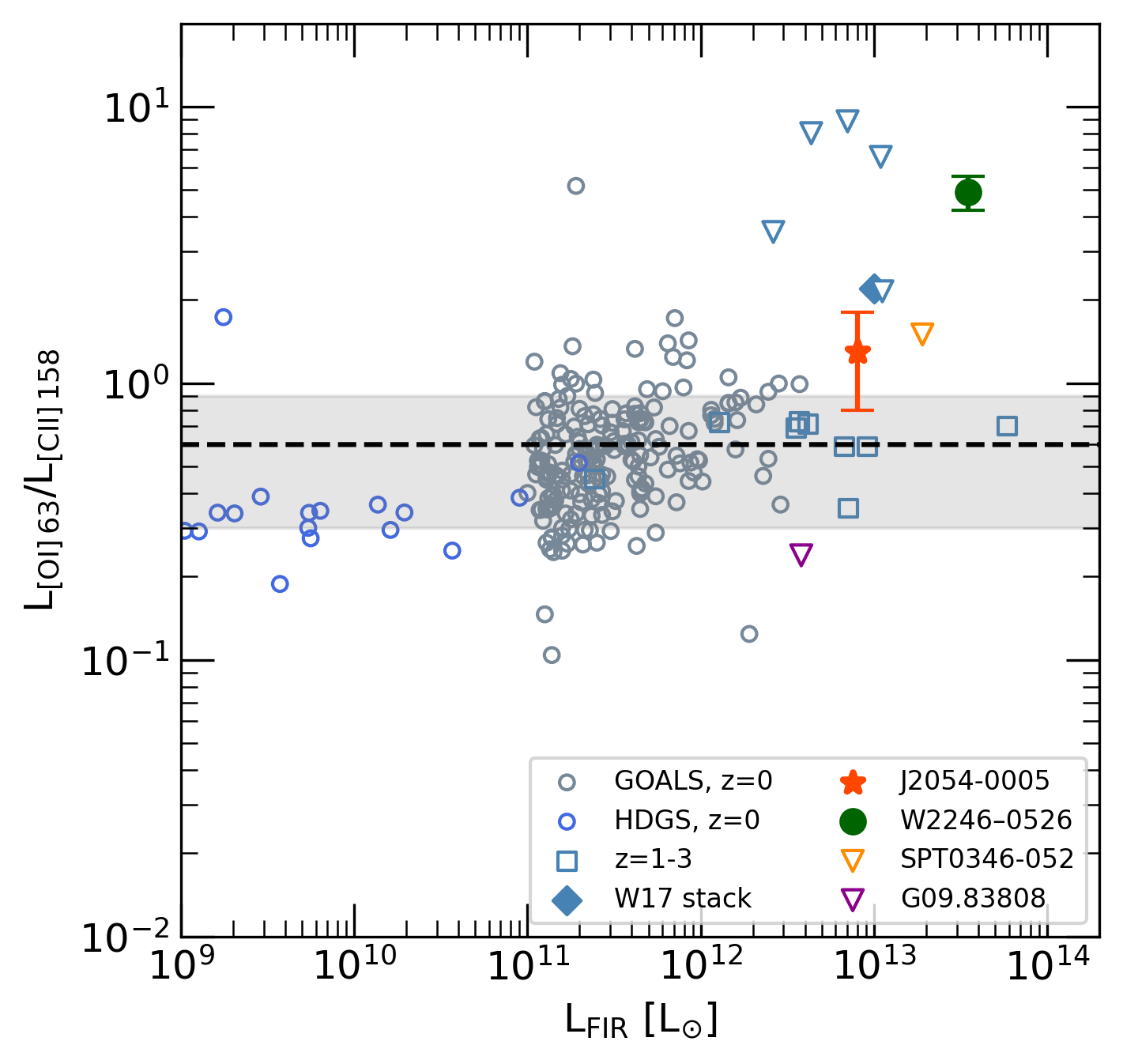}
\caption
{
{\bf (Left)}: The $L_{\rm [OI] 63 \micron}$-to-$L_{\rm FIR}$ ratio plotted against $L_{\rm FIR}$, where the luminosities are corrected for magnification, if any. The red star indicates the quasar J2054-0005 at $z = 6.04$. The other three objects at $z > 4$ are also shown: the detection of \oi\ 63 \micron\ in a hyper-luminous AGN W2246-0526 at $z=4.601$ (\citealt{fernandez_aranda2024}; green circle), and the non-detections of SMGs G09.83808 at at $z = 6.027$ (\citealt{rybak2023}; purple inverted triangle) and SPT 0346–52 at $z=5.656$ (\citealt{litke2022}; orange inverted triangle).
The plot includes nearby U/LIRGs from the GOALS survey (\citealt{diaz-santos2017}; small grey circles), blue compact dwarf galaxies from the Herschel Dwarf Galaxy survey (\citealt{cormier2015}; small blue circles), and other local galaxies observed by ISO (compiled in \citealt{coppin2012}; small black circles). A typical range of $L_{\mathrm{[OI]} 63 \, \mu\mathrm{m}}/L_{\mathrm{FIR}}$ in \cite{delooze2014} is shown by the horizontal dashed line with a shaded region. Also included are detections and non-detections of \oi\ 63 \micron\ at $z \sim 1-3$ (\citealt{coppin2012, brisbin2015, zhang2018, wagg2020}; blue squares, non-detections indicated by inverted triangles) and the detection of \oi\ 63 \micron\ in a stacked spectrum of four lensed SMGs (\citealt{wardlow2017}; blue diamond). The upper limits correspond to $3\sigma$. 
{\bf (Right)}: The $L_{\rm [OI] 63 \micron}$-to-$L_{\rm [CII] 158 \micron}$ ratio plotted against $L_{\rm FIR}$. The symbols are the same as in the left panel. The median and standard deviation of the local samples are shown as a horizontal black line with a shaded region. 
}
\label{fig2}
\end{figure*}

Table \ref{tab3} summarizes the previous line observations for J2054-0005\footnote{The line luminosities of \cii\ 158 \micron\ (\citealt{wang2013}) and \oiii\ 88 \micron\ (\citealt{hashimoto2019}) are obtained in the same manner as in \oi\ 63 \micron.}. 
We also list $L_{\rm FIR}$ obtained in \cite{salak2024} who have performed multi-wavelength spectral energy distribution (SED) fitting for J2054-0005 using {\tt CIGALE} (\citealt{boquien2019}) that includes the contribution from both star-formation and AGN activity. We obtain $L_{\rm [OI] 63 \micron}/L_{\rm FIR} = (6.7 \pm 2.2) \times 10^{-4}$ and $L_{\rm [OI] 63 \micron}/L_{\rm [CII] 158 \micron} = 1.3 \pm 0.5$. 

\subsubsection{$L_{\rm FIR}$ vs. $L_{\rm [OI] 63 \micron}/L_{\rm FIR}$}
\label{subsubsec:result_luminosityRatio1}

The left panel of figure \ref{fig2} shows a comparison of $L_{\rm [OI] 63 \micron}/L_{\rm FIR}$ with the literature values. 
These include nearby (U)LIRGs in the GOALS survey (\citealt{diaz-santos2017}; small grey circles), blue compact dwarf galaxies in the Herschel Dwarf Galaxy survey (\citealt{cormier2015}; small blue circles), and a compilation of other local galaxies observed by ISO (\citealt{coppin2012}; small black  circles\footnote{We have removed objects overlapped with \cite{diaz-santos2017}.}). \cite{delooze2014} have obtained an empirical relation between SFR and $L_{\rm [OI]63\micron}$ in the local universe. The horizontal line with a shaded region indicates the relation between $L_{\rm [OI] 63 \micron}/L_{\rm FIR}$ and $L_{\rm FIR}$, where we convert SFR into $L_{\rm FIR}$ under the assumption of the Salpeter initial mass function (\citealt{salpeter1955, kennicutt2012}).

The data also include the individual detections and non-detections of \oi\ 63 \micron\ at $z\sim 1-3$ (\citealt{coppin2012, brisbin2015, wardlow2017, zhang2018, wagg2020}; cyan squares), where the $3\sigma$ upper limits are shown with inverted triangles. The detection of \oi\ 63 \micron\ in a stacked spectrum of five lensed SMGs at $z\sim1-3$ is also shown (\citealt{wardlow2017}; cyan diamond).

We also include the data points of  $z>4$ ALMA observations, where the inverted triangle indicate the $3\sigma$ upper limits: detection in J2054-0005 (this study; red star), a $z=4.601$ hyper-luminous AGN W2246-0526 (\citealt{fernandez_aranda2024}; green circle), non-detection of two sub-millimeter galaxies (SMGs), G09.83808 at $z = 6.027$ (\citealt{rybak2023}; purple  inverted triangle) and SPT 0346–52 at $z=5.656$ (\citealt{litke2022}; orange inverted triangle). In this plot, we have corrected for the lensing magnification factors, if any. 

The left panel of figure \ref{fig2} shows that the data points of two \oi-detections at $z>4$ (J2054-0005 and W2246-0526) are well within the range of the local relation in \cite{delooze2014}, but are lower $L_{\rm [OI] 63 \micron}/L_{\rm FIR}$ than those in the individual detections at $z=1-3$.

As discussed in \cite{wardlow2017}, the initial campaign of \oi\ 63 \micron\ observations in high-$z$ SMGs in \cite{sturm2010} and \cite{coppin2012} have indicated that $L_{\rm [OI] 63 \micron}/L_{\rm FIR}$ may be high in the high-$z$ SMG population (cyan squares). On the other hand, the luminosity ratio in the stacked data of high-$z$ SMGs in \cite{wardlow2017} (cyan diamond) is consistent with the value obtained in the local universe. Interestingly, the three SMGs in \cite{coppin2012}\footnote{The object IDs are LESS66, 88, and 102.} and one SMG in \cite{sturm2010}\footnote{The object ID is MIPS J1248.} have AGN activity, which \cite{wardlow2017} have attributed to a possible explanation for the enhanced luminosity ratio. However, the situation is complicated as the sample of six SMGs at $z=1-2$ in \cite{brisbin2015} uniformly shows high $L_{\rm [OI] 63 \micron}/L_{\rm FIR}$ regardless of the presence of AGN activity, where three of the six objects have AGN activity (cyan squares). Based on the fact that $L_{\rm [OI] 63 \micron}/L_{\rm FIR}$ of the four objects at $z>4$ is consistent with that in the local universe, we conclude that high-$z$ SMGs do not necessarily have high $L_{\rm [OI] 63 \micron}/L_{\rm FIR}$.

\subsubsection{$L_{\rm FIR}$ vs. $L_{\rm [OI] 63 \micron\ }/L_{\rm [CII] 158 \micron}$}
\label{subsubsec:result_luminosityRatio2}

The right panel of figure \ref{fig2} shows a comparison of $L_{\rm [OI] 63 \micron}/L_{\rm [CII] 158 \micron}$ with the literature values. The symbols are the same as those in the left panel, where the data points of $z=1-3$ are plotted if the line ratio is available. 
The median and standard deviation of $L_{\rm [OI] 63 \micron}/L_{\rm [CII] 158 \micron}$ in the local samples are $0.6$ and $0.3$, respectively (horizontal black line and grey shaded region). 
The line ratios in the $z=1-3$ sample range from 0.5 to 2.0. The line raitos in the $z>4$ objects are $1.3\pm0.5$, $4.91\pm0.72$, $0.22$ ($3\sigma$) and $1.5$ ($3\sigma$) for J2054-0005, W2246-0526, G09.83808, and SPT 0346–52, respectively.

$L_{\rm [OI] 63 \micron}/L_{\rm [CII] 158 \micron}$ in J2054-0005 is consistent with the typical value obtained in the local universe within the uncertainty. On the other hand, W2246-0562 shows the second highest line ratio among the $z=0-6$ samples. 

\section{Discussion}
\label{sec:discussion}

\subsection{AGN contribution to \oi\ 63 \micron\ in J2054-0005}
\label{subsec:discussion1}

Previous studies have shown that $L_{\rm [OI] 63 \micron\ }/L_{\rm [CII] 158 \micron}$  is enhanced in the presence of AGN activity. Based on the observations of 52 objects in the Herschel SHINING survey, \cite{herrera-camus2018} have found that only objects with AGN activity reach $L_{\rm [OI] 63 \micron\ }/L_{\rm [CII] 158 \micron} \gtrsim 1.6$ in the local universe (see also \citealt{fukuchi2022}).

This is consistent with a theoretical study of \cite{abel2009} that predicts $L_{\rm [OI] 63 \micron\ }/L_{\rm [C II] 158 \micron}$ becomes higher in the presence of AGN activity. The calculations of \cite{abel2009} are based on CLOUDY (\citealt{ferland2013}) that include both star-formation and AGN activity, and adopt an one-dimensional spherical geometry. The authors have found that $L_{\rm [OI] 63 \micron\ }/L_{\rm [CII] 158 \micron} \gtrsim 2$ can be achieved only in the presence of AGN activity (see their figure 6). According to \cite{abel2009}, the higher line ratio is due to the fact that X-rays emitted by the AGN activity can penetrate deep into the cloud and heat the gas through photodissociatin and photoion-ization. Similarly, \cite{meijerink2007} have presented theoretical predictions for the $L_{\rm [OI] 63 \micron\ }/L_{\rm [C II] 158 \micron}$ ratio based on models accounting for both star formation and AGN activity. The result is consistent with \cite{abel2009} in the sense that the line ratio becomes higher in the presence of AGN activity. However, \cite{meijerink2007} have also shown that the line ratio can reach as high as $\sim 40$ with PDR models alone (i.e., without AGN acitivity) if the gas column density of PDR is sufficiently high (see their figure 2). Altogether, both the observational and theoretical studies show that $L_{\rm [OI] 63 \micron\ }/L_{\rm [CII] 158 \micron}$  is enhanced in the presence of AGN activity, although the exact value of the ratio is model-dependent.

In J2054-0005, the observed line ratio of $L_{\rm [OI] 63 \micron\ }/L_{\rm [CII] 158 \micron} = 1.3\pm0.5$ is close to the criterion presented in \cite{abel2009} and \cite{herrera-camus2018} within uncertainties, which does not allow us to strongly conclude whether the AGN activity affects the \oi\ 63 \micron\ line emission.

To further check if the AGN activity in J2054-0005 affects \oi\ 63 \micron, we refer to the results of \cite{decarli2020} who have concluded that PDRs rather than X-ray dominated regions (XDRs) better reproduce the observed lines of \cii\ 158 \micron, \ci(2-1), and CO in a sample of $z\sim6$ luminous quasars including J2054-0005. Briefly, the authors have constructed PDR and XDR models with a wide variety of parameter space in the hydrogen density ($n_{\rm H}$), incident radiation field, incident X-ray flux, and the total hydrogen column density ($N_{\rm HI}$) based on CLOUDY. The authors have found that the line ratio of $L_{\rm [CII] 158 \micron\ }/L_{\rm [CI](2-1)} > 10$ is achieved in PDRs, while it is $<10$ in XDRs at the column density of $N_{\rm HI}\gtrsim 10^{22}$ cm$^{-2}$ \footnote{This is due to the result of X-rays penetrate deeper into the clouds than UV photons, thus heating the cloud cores to enhance \ci(2-1) (\citealt{decarli2020}; see also \citealt{izumi2020} for the use of \cii/\ci\ to examine the presence of AGN).}. The observed line ratio of $L_{\rm [CII] 158 \micron\ }/L_{\rm [CI](2-1)} > 30$ indicates that the XDR does not strongly affect the emission lines in J2054-0005 (\citealt{decarli2020}). In conclusion, we do not find evidence in J2054-0005 that the emission lines are strongly powered due to the AGN activity.

\subsection{PDR Modeling}
\label{subsec:discussion2}

Under the assumption that the \oi\ 63 \micron\ and \cii\ 158 \micron\ lines are emitted not from XDRs but from PDRs in J2054-0005 (\S \ref{subsec:discussion1}), we examine the PDR properties of J2054-0005 based on comparisons of the observed line luminosities of \oi\ 63 \micron, \cii\ 158 \micron, and dust continuum emission (i.e., $L_{\rm FIR}$) to model predictions. These luminosities are commonly used to examine PDR properties (e.g., \citealt{hollenbach1999, malhotra2001}). The $L_{\rm [OI] 63 \micron\ }/L_{\rm [CII] 158 \micron}$ ratio is sensitive to $n_{\rm H}$ because the line ratio strongly depends on density beyond the critical density of \cii\ 158 \micron, $n_{\rm H, crit.} \sim 3 \times 10^{3}$~cm$^{-3}$, and \oi\ 63 \micron\ becomes an important coolant in the PDR at $G/G_{\rm 0}\gtrsim100$ (\citealt{wolfire2022}). The ratio of $(L_{\rm [OI] 63 \micron} + L_{\rm [CII] 158 \micron})/L_{\rm FIR}$ traces the heating efficiency (\citealt{wolfire2022}).

\begin{figure*}[thbp]
\hspace{+3cm}
\includegraphics[width=10cm]{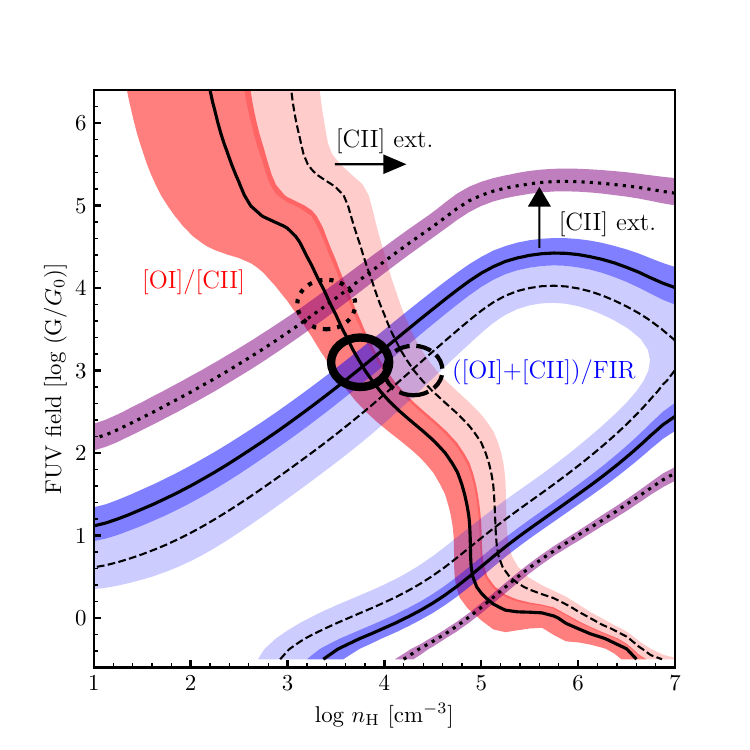}
\caption
{
The FUV radiation field, $G$, and the hydrogen gas density, $n_{\rm H}$, that reproduce the luminosity ratios of J2054-0005. We use the ``wk2020'' models of PDRToolbox (\citealt{pound_wolfire2023}). The darker red and blue shaded regions indicate the parameter space allowed by $L_{\rm [OI]63\micron}/L_{\rm [CII]158\micron}$ and $(L_{\rm [CII]158\micron} + L_{\rm [OI]63\micron})/L_{\rm FIR}$, respectively, where the black solid lines correspond to the median luminosity ratios. In this fiducial case, we have applied the following corrections to the luminosities (see \S \ref{subsubsec:discussion2_2} for the details); removal of the AGN contribution to $L_{\rm FIR}$, removal of the {\sc H ii} region contribution to $L_{\rm [CII]158\micron}$, and multiply $L_{\rm [OI]63\micron}$ by a factor of two. The fiducial result, log($n_{\rm H}$/cm$^{-3}$)$=3.7\pm0.3$ and log($G/G_{\rm 0}$)~$= 3.0\pm0.1$, is shown by a thick black circle. 
The lighter red and blue shaded regions indicate the parameter space if \oi\ 63 \micron\ is self-absorbed by a factor of two, i.e., $L_{\rm [OI]63\micron}$ multiplied by an additional factor of two (see \S\ref{subsubsec:discussion2_3}). The black dashed lines correspond to the median luminosity ratios. The result, log($n_{\rm H}$/cm$^{-3}$)$=4.3\pm0.3$ and log($G/G_{\rm 0}$)~$= 3.0\pm0.4$, is shown by a black dashed circle.
As a reference, the purple shaded region and the dotted line indicate the parameter space if no correction is applied to $L_{\rm FIR}$. The result, log($n_{\rm H}$/cm$^{-3}$)$=3.4\pm0.4$ and log($G/G_{\rm 0}$)~$= 3.7\pm0.3$, is shown by a black dotted circle. 
Two black arrows indicate the direction of correction to account for the different spatial size of \cii\ 158 \micron\ and dust continuum emission.
}
\label{fig3}
\end{figure*}

\subsubsection{Model Descriptions}
\label{subsubsec:discussion2_1}

We employ PDR Toolbox (PDRT; \citealt{pound_wolfire2008, pound_wolfire2023}). The models have two codes, the ``Wolfire-Kauffman" code and the KOSMA-tau code (\citealt{rollig2013, pound_wolfire2023}). Following recent observational studies of high-$z$ quasars (e.g., \citealt{shao2019, yang2019}), we use the latest version of the ``Wolfire-Kauffman" code, which is referred to as the ``wk2020" model in \cite{pound_wolfire2023}. The model assumptions are summarized in table 1 in \cite{pound_wolfire2023}. Briefly, the model assumes a plane-parallel geometry with UV radiation field, cosmic rays, and soft X-rays incident on one side. The model also assumes abundances of e.g., carbon, oxygen, and polycyclic aromatic hydrocarbons (PAHs). The ``wk2020'' models have two sets of metallicity, $Z/Z_{\odot} = 0.5$ and $1.0$. In this study, we adopt the case of $Z/Z_{\odot}=1.0$\footnote{Although the metallicity is not obtained for J2054-0005 in the literature due to the lack of lines necessary to estimate it, the choice of the value is motivated by the metallicity estimates in two similarly FIR-luminous quasars: \cite{li2020} and \cite{novak2019} have obtained $Z/Z_{\odot} = 1.5 - 2.1$ and $0.7 - 2.0$ for J2310+1855 at $z=6.0$ and J1342+0928 at $z=7.54$, respectively.}.

The two main input parameters are the radiation field strength in terms of Habing fields ($G_{\rm 0} = 1.6 \times 10^{-3}$~erg~cm$^{-2}$~s$^{-1}$) and a constant hydrogen nucleus density, $n_{\rm H}$. With these two parameters, the model calculates local chemical equilibrium to determine local density, solves local energy balance to estimate temperature and pressure, and performs radiative transfer through a PDR layer to output line intensities from one side. Thus, if more than two intensity ratios are available, one can constrain $n_{\rm H}$ and the FUV radiation field based on comparisons of the models with observed line intensities. Note that PDRT assumes that \cii\ 158 \micron\ is purely emitted from the PDR. In this study, we further assume that \cii, \oi, and dust continuum emission are co-spatial, but discuss this limitation in \S \ref{subsubsec:discussion2_3}.

\subsubsection{Corrections to Input Luminosities}
\label{subsubsec:discussion2_2}

Before comparing the models to the line luminosity ratios in J2054-0005, we apply the following three corrections to the data. First, although we assume that the \oi\ 63 \micron\ and \cii\ 158 \micron\ lines are emitted from PDRs (\S \ref{subsec:discussion1}), it is possible that the dust continuum emission is partly powered by the AGN activity in J2054-0005. 
Based on the results of multi-wavelength SED fitting with {\tt CIGALE} in \cite{salak2024}, we find that the AGN activity contributes to $\approx 78$\%\ of the FIR luminosity on the galactic scale in J2054-0005\footnote{In \cite{salak2024}, the fractional AGN contribution to the IR luminosity is computed based on \cite{fritz2006} that assumes three components through a radiative transfer: primary source located in the torus, the scattered emission by dust, and the thermal dust emission.}.
We thus adopt the far-infrared luminosity from star formation activity, $L_{\rm FIR}$(PDR) $= 0.22~L_{\rm FIR} = (1.5\pm0.1)\times10^{12}~L_{\rm \odot}$.

Secondly, although the PDRT models assume that \cii\ 158 \micron\ is purely emitted from PDRs, \cii\ originates from both the {\sc Hii} regions and PDR in reality (e.g., \citealt{Oberst2006, herrera-camus2016}). We assume that $17$\%\ of \cii\ 158 \micron\ is emitted from the {\sc Hii} regions based on a result of a similarly FIR-luminous quasar, J2310+1855 at $z=6.0$\footnote{The contribution from the {\sc Hii} regions can be estimated with the luminosity ratio of \cii-to-\nii\ 205 \micron. However, \nii\ 205 \micron\ is not observed in J2054-0005.} (\citealt{li2020}). The \cii\ 158 \micron\ luminosity emitted from PDR, $L_{\rm [CII] 158 \micron}$(PDR), is estimated to be $0.83~L_{\rm [CII] 158 \micron} = (2.8\pm0.4)\times10^{9}~L_{\rm \odot}$.

Finally, in a more realistic geometry of spherical clouds, the optically thin emission (FIR continuum and \cii\ 158 \micron) would be detected from both the front and back sides of the cloud, whereas the optically thick emission (\oi\ 63 \micron) would be detected only from the front side\footnote{\cii\ 158 \micron\ becomes optically thick at the column density of $N$(C$^{+}$)~$=$~$5\times10^{17}$~cm$^{-2}$ at a velocity width of 4~km~s$^{-1}$ (\citealt{russell1980}). On the other hand, the optical depth of \oi\ 63 \micron\ becomes unity at a column density of $N$(O$^{0}$)~$=$~$2\times10^{17}$~cm$^{-2}$ at a velocity width of 1~km~s$^{-1}$ (\citealt{wolfire2022}). Because O$^{0}$ is present up to higher optical depth clouds than C$^{+}$, \oi\ 63 \micron\ becomes opitcally-thick faster than \cii\ 158 \micron\ (\citealt{malhotra2001}).} (\citealt{kaufman1999, rollig2007, yang2019}). Following previous studies (e.g., \citealt{wardlow2017, yang2019, hashimoto2023}), we multiply the observed luminosity of \oi\ 63 \micron\ by a factor of two, $L_{\rm [OI] 63\micron, corr.} = 2~L_{\rm [OI] 63\micron} = (9.1\pm3.0)\times10^{9}~L_{\rm \odot}$.

\subsubsection{Results of PDR Modeling}
\label{subsubsec:discussion2_3}

Figure \ref{fig3} shows the results of PDR modeling in J2054-0005. The darker red and blue shaded regions indicate the parameter space allowed by $L_{\rm [OI]63\micron}/L_{\rm [CII]158\micron}$ and $(L_{\rm [CII]158\micron} + L_{\rm [OI]63\micron})/L_{\rm FIR}$, respectively. We obtain a fiducial result, log($n_{\rm H}$/cm$^{-3}$)$=3.7\pm0.3$ and log($G/G_{\rm 0}$)~$= 3.0\pm0.1$, as indicated by a thick black circle. Here, we only consider the solution with log($G/G_{\rm 0})$~$\gtrsim 1$ as physically plausible following the discussion in \cite{brisbin2015}.

Because the correction to $L_{\rm FIR}$ in \S \ref{subsubsec:discussion2_2} is not usually applied even in the presence of AGN activity (e.g., \citealt{brisbin2015}), as a comparison, we also perform PDR modeling with the same models without applying the correction to $L_{\rm FIR}$ (purple shaded region). As indicated by a black dotted circle, the result, log($n_{\rm H}$/cm$^{-3}$)$=3.4\pm0.4$ and log($G/G_{\rm 0}$)~$= 3.7\pm0.3$, is shifted toward a higher FUV radiation field compared to the fiducial case. Given the possible uncertainty in the correction of $L_{\rm FIR}$, the FUV radiation field of log($G/G_{\rm 0}$)~$\sim 3.7$ can be regarded as the upper limit.

\begin{figure}[t]
\vspace{-0.2cm}
\includegraphics[width=9cm]{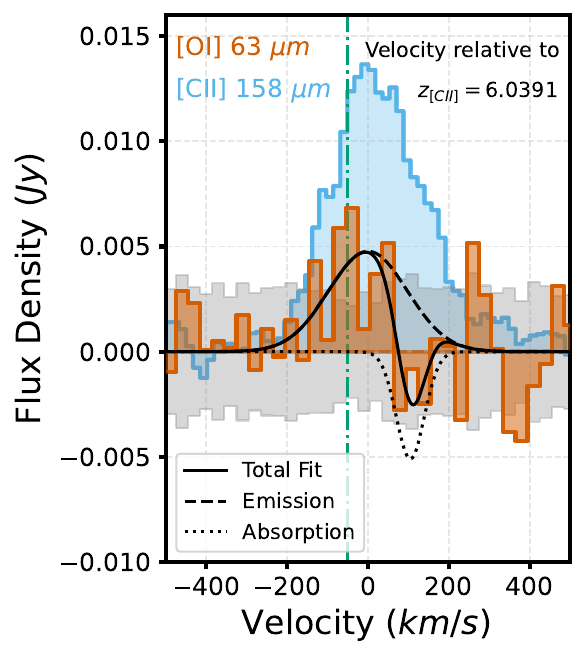}
\caption
{The \oi\ 63 \micron\ (orange) with error (grey) plotted over the \cii\ line (blue) from \cite{wang2013}. The \oi\ 63 \micron\ line, if it is not self-absorbed, appears narrow and has a line center (green dash-dotted) that is $\sim$~50~km~s$^{-1}$ blueward of the \cii\ line, suggesting different spatial extent of the to emission lines in J2054-0005. If we assume that the both lines arise in the same regions, consistent with our PDR modeling, then the \oi\ spectrum is well described via self-absorption with a two component gaussian model (solid black) including an emission (dashed black) and absorption (dotted black) component. 
}
\label{fig4}
\end{figure}

We further discuss two considerations in the PDR modeling. Firstly, the \oi\ 63 \micron\ line may be self-absorbed as described above. Given the limited quality of our data, it is difficult to determine conclusively if the line in J2054-0005 is indeed self-absorbed. However, by closely comparing the \oi\ 63 \micron\ line profile to that of \cii\ 158 \micron\ as shown in figure \ref{fig4}, we can make a tentative assessment. In this figure, we show the spectrum at a finer resolution of 30~km~s$^{-1}$. Examining the two lines in figure \ref{fig4}, if we assume that the \oi\ 63 \micron\ line is not self-absorbed, then its line peak is $\sim 50$~km~s$^{-1}$ blueward of the \cii\ peak, with a line width about half as narrow as the \cii\ line. While this scenario is not impossible, especially if the \cii\ line is more spatially extended, it may also be hinting that some of the \oi\ line has been self-absorbed. To assess how much, a two component-gaussian fit is made to the \oi\ data. For the fit we assume that the \oi\ emission should arise in the same regions as the \cii\ so we force one component to the same line-center and line-width as the \cii\ line, letting only its amplitude vary. The second component is left completely unconstrained. The result of the non-linear least-squares fit via {\it Scipy}’s {\tt ODR} module is shown in figure \ref{fig4}. The best fit model includes an absorption component at $\sim 100$~km~s$^{-1}$ with a FWHM of $\sim 80$~km~s$^{-1}$. Integrating the emission component we get an estimated line-flux of $2.9 \pm 0.7$ Jy km~s$^{-1}$, about 50\% higher than if we assume the line is not self-absorbed. Given the low signal-to-noise ratio of the spectrum, this exact value should be treated cautiously, but it does suggest that the \oi\ 63 \micron\ line in J2054-0005 may indeed be self-absorbed. To account for this possibility, we also perform PDR modeling by multiplying the \oi\ 63 \micron\ line luminosity by an additional factor of two\footnote{As discussed in \cite{wolfire2022}, the correction to \oi\ 63 \micron\ in the case of self-absorption is typically $2-4$. This is also in agreement, within the uncertainties, with the un-absorbed flux estimated via our gaussian fitting.} (lighter red and blue shaded regions in figure \ref{fig3}).
As indicated by a black dashed circle, the result, log($n_{\rm H}$/cm$^{-3}$)~$=4.3\pm0.3$ and log($G/G_{\rm 0}$)~$= 3.0\pm0.4$, is slightly shifted toward a higher gas density.

Secondly, the \cii\ 158 \micron, \oi\ 63 \micron, and dust continuum emission are not necessarily co-spatial. Deep ALMA imaging has revealed extended, diffuse \cii-``halos'' around high-$z$ star-forming galaxies (e.g., \citealt{gullberg2018, fujimoto2019, fujimoto2020, rybak2019, rybak2020, ikeda2024}). Recently, based on the high-angular resolution data of \cii\ 158 \micron\ in the ALMA-CRISTAL survey, \cite{ikeda2024} have found that the spatial extent of \cii\ can be explained by PDRs, while the contribution from diffuse neutral medium (atomic gas) and the effects of mergers may further expand the \cii\ line distributions. In case of J2054-0005, high-resolution ALMA imaging indicates that the \cii\ emission is a factor of $\approx 2$ more extended than dust (\citealt{venemans2020}). In figure \ref{fig3}, we show an arrow indicating the direction of the correction due to the extended \cii, where we divide the \cii\ luminosity by a factor of two. A detailed discussion is hampered by the fact that the spatial extent of \oi\ is not known. Future higher-resolution observation of \oi\ 63 \micron\ is crucial for more sophisticated PDR modeling.

\subsubsection{Comparisons of Our PDR Modeling to the Literatures}
\label{subsubsec:discussion2_4}

\begin{figure}[t]
\vspace{-0.2cm}
\includegraphics[width=9cm]{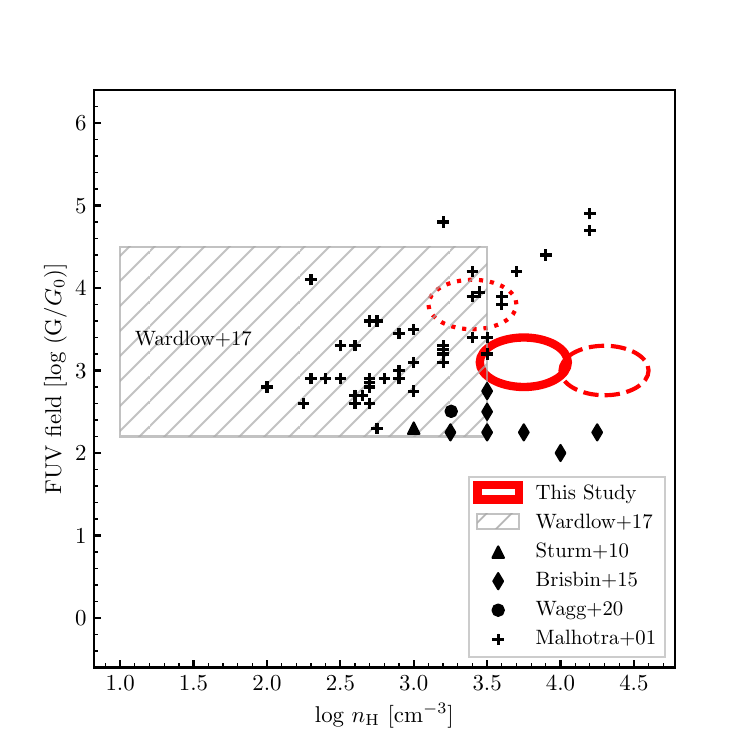}
\caption
{
Comparisons of the FUV radiation field, $G$, and the hydrogen gas density, $n_{\rm H}$, obtained in J2054-0005 with the literature values. In J2054-0005, three results are shown with red circles (see figure \ref{fig3}). 
The black crosses show the results of nearby star-forming galaxies (\citealt{malhotra2001}). The black triangle and diamonds indicate the results of a lensed $z\sim1$ SMG (\citealt{sturm2010}) and unlensed SMGs at $z\sim1-2$ (\citealt{brisbin2015}). The black circle shows a $z\sim1$ BzK galaxy (\citealt{wagg2020}). The region with a hatch corresponds to the result of a stacked spectrum of $z\sim1-3$ SMGs in \cite{wardlow2017}. J2054-0005 has the PDR properties broadly consistent with those in the literatures. 
}
\label{fig5}
\end{figure}

In figure \ref{fig5}, we compare the results of J2054-0005 with the literature values. Black crosses and a black triangle show the results in nearby star-forming galaxies (\citealt{malhotra2001}) and a lensed $z\sim1$ SMG (\citealt{sturm2010}), respectively. These results are obtained based on the models of \cite{kaufman1999} with the \oi\ 63 \micron, \cii\ 158 \micron, and IR luminosities. 
Black diamonds indicate the results in non-lensed SMGs at $z\sim1-2$ (\citealt{brisbin2015}). The authors have used the PDR models of \cite{pound_wolfire2008} with the \oi\ 63 \micron, \cii\ 158 \micron, and IR luminosities. 
A black circle shows the result of \cite{wagg2020} in a BzK galaxy at $z\sim1$. The PDR modeling is based on the models of \cite{wolfire2022} with the \oi\ 63 \micron, CO(2-1), and IR luminosities. 
Finally, we also plot the result of \cite{wardlow2017} who have performed PDR modeling for a stacked spectrum of $z\sim1-3$ SMGs. The authors have used the models of \cite{pound_wolfire2008} and  \oi\ 63 \micron, \cii\ 158 \micron, \siiii\ 34 \micron, and IR luminosities.

As can be seen in figure \ref{fig5}, the samples of nearby star-forming galaxies and $z=1-3$ SMGs with individual detections of \oi\ 63 \micron\ exhibit a broad range of gas density, ranging from log($n_{\rm H}$/cm$^{-3}$) $\sim~2$ to $>~4$. The PDR properties of J2054-0005 are broadly consistent with these galaxies, however, we note that J2054-0005 falls within the region that overlaps with the high log($n_{\rm H}$) objects among the star-forming galaxies. This is consistent with a picture that the gas is centrally concentrated in the quasar host galaxy, leading to the stimulation of both intense nuclear star formation and central supermassive black hole accretion (e.g., \citealt{hopkins2008}).

\section{Summary}
\label{sec:summary}

We have presented \oi\ 63 \micron\ data of a luminous quasar, J2054-0005 at $z=6.04$ obtained with ALMA Band 9. The \oi\ 63 \micron\ line is detected at the peak significance level of $5.0$ at the expected position and frequency (figure \ref{fig1}). To date, this is the highest-$z$ detection of \oi\ 63 \micron, and only the second case of \oi\ 63 \micron\ detection at $z>4$ after a hyper-luminous AGN, W2246-0526 (\citealt{fernandez_aranda2024}). Our findings are summarized as follows. 

\begin{itemize}
\item In J2054-0005, the \oi\ 63 \micron\ line luminosity is $(4.5\pm1.5)~\times~10^{9}~L_{\rm \odot}$, corresponding to $L_{\rm [OI] 63 \micron}/L_{\rm FIR}~=~(6.7\pm2.2)\times10^{-4}$. We have compiled the data points of objects with \oi\ 63 \micron\ observations at $z=0-4$, and compared them with J2054-0005. In contrast to the previous \oi\ 63 \micron\ detections in individual SMGs at $z=1-3$ that show enhanced $L_{\rm [OI] 63 \micron}/L_{\rm FIR}$ against the local value, we have found that the values in J2054-0005 and W2246-0526 at $z=4-6$ are consistent with those in the local universe (left panel of figure \ref{fig2}). In J2054-0005, \oi\ 63 \micron\ is as bright as \cii\ 158 \micron, resulting in $L_{\rm [OI]63\micron}/L_{\rm [CII]158\micron} = 1.3\pm0.5$ (right panel of figure \ref{fig2}).
\item Analyzing the \oi\ 63 \micron\ spectrum and fitting a two-component Gaussian model suggests that the \oi\ 63 \micron\ line may be self-absorbed. Deeper observations of the line, to significantly detect the absorption feature are necessary to conclusively determine the degree of self-absorption (figure \ref{fig4}).
\item We have performed PDR modeling with PDRT (\citealt{pound_wolfire2023}) based on the luminosity ratios of $L_{\rm [OI] 63 \micron\ }/L_{\rm [CII] 158 \micron}$ and $(L_{\rm [OI] 63 \micron} + L_{\rm [CII] 158 \micron})/L_{\rm FIR}$. We have carefully removed the possible contribution of AGN activity to $L_{\rm FIR}$ and {\sc H ii} contribution to $L_{\rm [CII]158\micron}$. We have obtained the gas density of log($n_{\rm H}$/cm$^{-3}$)$=3.7\pm0.3$ and FUV radiation field of log($G/G_{\rm 0}$)~$= 3.0\pm0.1$, although the results can slightly change if we consider the self-absorption effect on \oi\ 63 \micron. The PDR properties are in a broad agreement with those obtained in nearby star-forming galaxies and other $z=1-3$ SMGs with individual \oi\ 63 \micron\ detections (figures  \ref{fig3} and \ref{fig5}). We note that J2054-0005 falls within the region that overlaps with the high log($n_{\rm H}$) objects among the star-forming galaxies.
\end{itemize}

In this work, we have demonstrated the power of \oi\ 63 \micron\ to probe the dense neutral gas in high-$z$ galaxies. Although \oi\ 63 \micron\ is not easy to observe in $z\sim4-7$ galaxies because the line is redshifted in high-frequency bands of ALMA (Band 9 and 10), a statistical sample of \oi\ 63 \micron\ observations is crucial to study the PDR gas in quasar host galaxies and compare them with those in normal star-forming galaxies at similar epochs.

\section*{Acknowledgments}
The authors thank the anonymous referee for comments and suggestions that helped us improve the manuscript. We also acknowledge Ken Ohsuga, Ryota Ikeda, Atsushi Yasuda, Saho Kawahara, Ryota Ura, Tatsuya Matsumura, Yu Nagai for useful discussion. 
N.I. was supported by the ALMA Japan Research Grant of NAOJ ALMA Project, NAOJ-ALMA-341.
T.H. was supported by Leading Initiative for Excellent Young Researchers, MEXT, Japan (HJH02007) and by JSPS KAKENHI grant Nos. 20K22358 and 22H01258. M.R. was supported by the NWO Veni project ``\textit{Under the lens}" (VI.Veni.202.225). T.M. was supported by a University Research Support Grant from the National Astronomical Observatory of Japan (NAOJ). D.D acknowledges support from the National Science Center (NCN) grant SONATA (UMO-2020/39/D/ST9/00720). K.M. acknowledges support from the JapanSociety for the Promotion of Science (JSPS) through KAKENHI grant No. 20K14516. Y.T. was supported by KAKENHI grant Nos. 22H04939 and 23K20035. T.I. was supported by KAKENHI No. 23K03462. Y.N. acknowledges funding from JSPS KAKENHI Grant Number 23KJ0728. 
This paper makes use of the following ALMA data: ADS/JAO.ALMA\#2016.1.01063.S. ALMA is a partnership of ESO (representing its member states), NSF (USA) and NINS (Japan), together with NRC (Canada), MOST and ASIAA (Taiwan), and KASI (Republic of Korea), in cooperation with the Republic of Chile. The Joint ALMA Observatory is operated by ESO, AUI/NRAO and NAOJ.


\end{document}